\documentclass[conference]{IEEEtran}
\ifCLASSINFOpdf
\else
\fi
		\usepackage{enumerate}
	\usepackage{bbding}
	\usepackage{makecell}
	\usepackage{ragged2e} 
	\usepackage{booktabs,makecell, multirow, tabularx}
	\usepackage{mathtools} 
	\usepackage{graphicx}
	\usepackage{amsmath}
	\usepackage{graphicx, subfig}
	\usepackage{amssymb}
	\makeatletter
	
	\newcommand{\Rmnum}[1]{\expandafter\@slowromancap\romannumeral #1@}
	\makeatother
	
	\usepackage{algorithmic}
	\usepackage{amsfonts}
	\usepackage{url}
	\usepackage[T1]{fontenc} 
	\usepackage[misc]{ifsym}

\begin{document}
		%
		\title{A Key-Driven Framework for Identity-Preserving Face Anonymization}

		
		
		%
	
		\author{\IEEEauthorblockN{Miaomiao Wang\IEEEauthorrefmark{2},
				Guang Hua\IEEEauthorrefmark{3},
				Sheng Li\IEEEauthorrefmark{4}$ ^{\textrm{\Letter}}$, and
				Guorui Feng\IEEEauthorrefmark{2}$^{\textrm{\Letter}}$ }
			\IEEEauthorblockA{\IEEEauthorrefmark{2}School of Communication and Information Engineering,  Shanghai University, China }
			\IEEEauthorblockA{\IEEEauthorrefmark{3}Infocomm Technology Cluster, Singapore Institute of Technology, Singapore}
			\IEEEauthorblockA{\IEEEauthorrefmark{4}School of Computer Science, Fudan University, China}
	     	}
	
		
		\IEEEoverridecommandlockouts
		\makeatletter\def\@IEEEpubidpullup{6.5\baselineskip}\makeatother
		\IEEEpubid{\parbox{\columnwidth}{\rule{\columnwidth/2}{0.5pt}\\
			     $^{\textrm{\Letter}}$ Corresponding authors: Sheng Li <lisheng@fudan.edu.cn>, and Guorui Feng <grfeng@shu.edu.cn>\\ \\
				Network and Distributed System Security (NDSS) Symposium 2025\\
				23 - 28 February  2025, San Diego, CA, USA\\
				ISBN 979-8-9894372-8-3\\
				https://dx.doi.org/10.14722/ndss.2025.23729\\
				www.ndss-symposium.org
			}
			\hspace{\columnsep}\makebox[\columnwidth]{}}
		
		
		\maketitle
		
		\begin{abstract}
			
			Virtual faces are crucial content in the metaverse. Recently, attempts have been made to generate virtual faces for privacy protection. Nevertheless, these virtual faces either permanently remove the identifiable information or map the original identity into a virtual one, which loses the original identity forever. In this study, we first attempt to address the conflict between privacy and identifiability in virtual faces, where a  key-driven face anonymization and authentication recognition (KFAAR) framework is proposed. Concretely, the KFAAR framework consists of a head posture-preserving virtual face generation (HPVFG) module and a key-controllable virtual face authentication (KVFA) module. The HPVFG module uses a user key to project the latent vector of the original face into a virtual one. Then it maps the virtual vectors to obtain an extended encoding, based on which the virtual face is generated. By simultaneously adding a head posture and facial expression correction module, the virtual face has the same head posture and facial expression as the original face. During the authentication, we propose a KVFA module to directly recognize the virtual faces using the correct user key, which can obtain the original identity without exposing the original face image. We also propose a multi-task learning objective to train HPVFG and KVFA. Extensive experiments demonstrate the advantages of the proposed HPVFG and KVFA modules, which effectively achieve both facial anonymity and identifiability.
		\end{abstract}
		
		
		%
		
		\setcounter{page}{1}
		
		\section{Introduction}
		\subsection{Background}
		
		\begin{figure}[!t]
			\centering 
			\includegraphics[width=\linewidth]{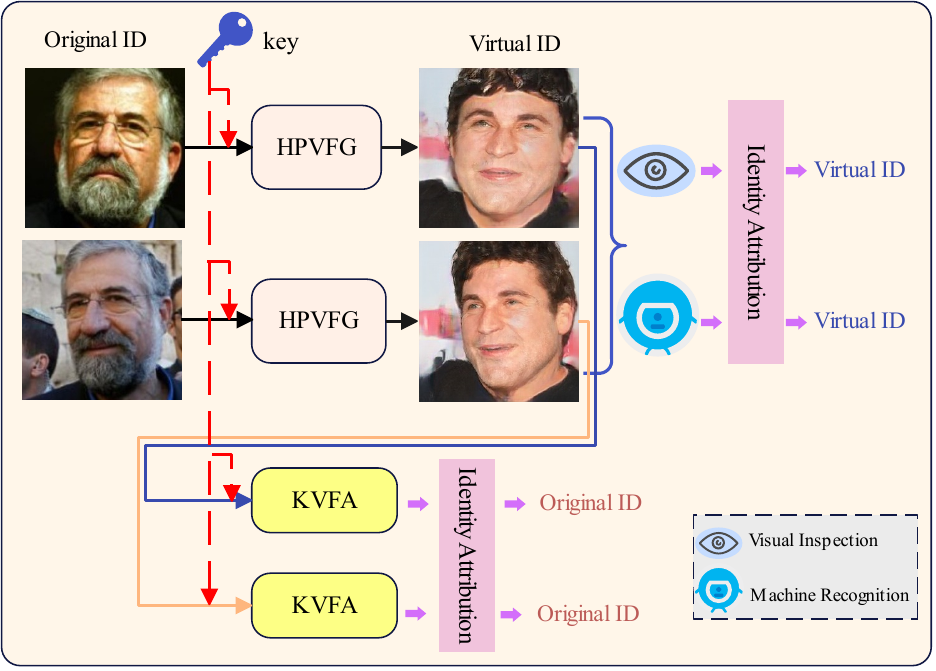}
			\caption{Illustration of the proposed framework. The virtual face generated by our HPVFG has significant differences in terms of visual appearance and machine recognition from the original face. With the correct key, it can be authenticated through KVFA to match their original identity.}
			\label{fig_1}    
		\end{figure}
		
		The virtual face is widely used in the metaverse, which can be generated by artificial intelligence content generation techniques. Research on virtual faces provides more possibilities for the development of the metaverse. 
		However, the widespread application of virtual faces has also brought about a series of issues regarding privacy, ethics, and security. Therefore, when discussing the application of virtual faces in the metaverse, we also need to seriously think about how to balance the relationship between technological innovation and the protection of personal rights and interests.  In addition, to meet the needs of interactivity, virtual faces need to be able to respond to user actions and emotions in real time. 
		Early approaches protect the privacy of faces through masking techniques such as mosaic \cite{ref38}, blurring \cite{ref1} and pixelization \cite{ref2}. These methods can anonymize digital face images, but the virtual face images often have poor visual quality, which hinders subsequent applications \cite{ref3}.
		With the recent advances in deep learning, researchers take advantage of generative adversarial networks \cite{ref4} to generate virtual faces \cite {ref5, ref36, ref6, ref7}. Despite the high quality and anonymization ability, these face images suffer from permanent identity loss and cannot be utilized for recognition.
		To tackle the privacy-utility trade-off, recent efforts \cite {ref8, ref9, ref11, ref12} have been made to generate the visual faces while maintaining the identifiability.
		
		Li et al. \cite{ref8} propose a method that adaptively locates identity-independent attributes in face images and generates the virtual faces using the original face and the located face attributes. They subsequently \cite{ref9} propose an identity-preserving face disguise method by decoupling the appearance code and the identification code, and replacing the appearance code with a target one. Although these methods can ensure the identifiability of the virtual faces, the identity information is still revealed, posing privacy risks. The work in \cite{ref12} proposes an identifiable virtual face generator (IVFG), which does not reveal any information about the original face during the authentication. Specifically, it uses different keys to control the identities of the virtual faces, which can be directly used for face authentication. However, the virtual faces cannot meet the requirements of posture synchronization, which is always the same regardless of the head posture of the original face images. In addition, face recognition is conducted in the virtual domain with the original identity lost.

		\begin{table*}[t]
			\caption{Comparison Between Our Method and Mainstream Methods}
			\center
			\label{tab:1}
			\begin{tabular}{cccccccccc}
				\toprule
				
				Methods& Realism  & Diversity &Synchronism& Virtual Background&Virtual Look&Virtual ID& Head Posture&Identity Authentication &Key Policy\\
				\midrule
				CIAGAN \cite{ref6} &\XSolidBrush &\XSolidBrush & \XSolidBrush&\XSolidBrush& \CheckmarkBold &\CheckmarkBold & \CheckmarkBold&\XSolidBrush&\XSolidBrush\\
				VFGM \cite{ref37}  &\CheckmarkBold &\CheckmarkBold & \CheckmarkBold&\CheckmarkBold& \CheckmarkBold &\XSolidBrush & \CheckmarkBold&\CheckmarkBold&\XSolidBrush\\
				IVFG \cite{ref12} &\CheckmarkBold &\CheckmarkBold & \CheckmarkBold&\CheckmarkBold& \CheckmarkBold &\CheckmarkBold & \XSolidBrush&\XSolidBrush&\CheckmarkBold\\				
				Ours &\CheckmarkBold &\CheckmarkBold & \CheckmarkBold&\CheckmarkBold& \CheckmarkBold &\CheckmarkBold & \CheckmarkBold&\CheckmarkBold&\CheckmarkBold\\
				
				\bottomrule
			\end{tabular}
		\end{table*}

		In recent years, methods focused on the interactivity of virtual faces have been proposed. The generator of CIAGAN \cite{ref6} is able to preserve the head posture after the anonymization. However, the stability of the image quality still faces certain challenges, where the generated images may be blurred, distorted, or have artifacts. In addition, the original identity of the virtual face is permanently lost and cannot be applied to downstream tasks like face recognition and authentication. Since the latent space of the StyleGAN2 contains rich semantic information, VFGM \cite{ref37} performs style mixing in the latent space to generate virtual faces that maintain head posture. However, this method only provides privacy protection at the visual level, which does not offer any protection in terms of face recognition. That is to say, the protected face image is visually different from the original face, but the attacks can still obtain the original identity using a face recognition approach.
		
		To summarize, existing research on facial anonymization faces the following challenges. \textbf{\textit{(1) The conflict between anonymity and identifiability.}} The existing virtual faces either completely discard the identifiability or transform the original identity into a virtual one. Both permanently lose their original identity, which creates obstacles for face management. \textbf{\textit{(2) Difficulties in maintaining head posture and facial expression.}} The virtual faces need to maintain a certain level of authenticity with the head posture and facial expression synchronized for face recognition tasks such as head turning, smiling, and blinking, to enhance the interaction between the real world and the virtual world.
		
		
		\subsection{Our Work}
		In order to solve the above problems, we aim to propose an identity-preserving framework with the following functionalities, (1) authorized and traceable virtual face authentication to balance between privacy and authenticity. It is required that the virtual face can be authenticated as the original identity with the correct key, and no visual information is leaked during the authentication.
		And (2) significant appearance changes with the head posture maintained. In the metaverse, the virtual face has significant visual differences from the original face while maintaining the head posture and expression characteristics of the original face. These are achieved through a key-driven face anonymization and authentication recognition (KFAAR) multitasking mechanism, which contains a head posture-preserving virtual face generation (HPVFG) module and a key controllable virtual face authentication (KVFA) module.
		
		Our HPVFG module leverages a user-specific key to project the latent vector of the original face into a virtual vector space, ensuring that the virtual face exhibits significant visual differences from the original face while maintaining the original head posture and expression. This is accomplished through a series of components, including an encoder, a projector, a mapping network, a generator, and a head posture correction module. These components jointly produce a virtual face that is anonymized with the original head posture and facial expression maintained.
		
		The KVFA module, on the other hand, is designed for virtual face authentication. It is capable of extracting the original identity from the virtual face using a dedicated recognition network when the correct key is given. This network is trained through a multi-task learning objective, which can prevent misidentification and false rejection, and guarantees correct face recognition without revealing the visual information of the original face.
		
		The flowchart of the proposed framework is presented in Figure \ref{fig_1}. It can be seen that the virtual faces generated by our HPVFG have significant visual differences from the original face. Therefore neither human nor machine can determine its true identity. It should also be noted that our virtual faces can maintain the original head postures and facial expression. With the correct key given, face authentication can be performed through KVFA, allowing the virtual face to be identified as the original identity. Table \ref{tab:1} summarizes the attributes of our virtual faces in comparison with the mainstream solutions, including CIAGAN \cite{ref6}, VFGM \cite{ref37} and IVFG \cite{ref12}.
		
		We summarize the main contributions as follows:
		\begin{itemize}
			\item {We make the first attempt to solve the virtual face authentication challenge while offering privacy protection. The generated high-quality virtual faces are anonymized to both humans and machines, while the original identity can still be obtained with the correct key for authentication.} 
			\item{We propose a new framework HPVFG to generate anonymous, synchronized, diverse, and high-quality virtual faces.}
			\item{We propose a dedicated recognition network i.e., the KVFA module, which can extract the original identities from our virtual faces using the correct key.}
			\item{Extensive experiments demonstrate the effectiveness of our proposed framework. Compared with SOTA, our method performs well in generating the virtual faces that maintain head posture and facial expression, which achieves face  authentication through the KVFA module. In addition, we also conduct a security analysis of the potential attacks to demonstrate the security of our method.}
			
		\end{itemize}

		\section{Related Work}
		We classify the face anonymization methods into the following three categories.
		
		\textbf{Anonymization by Visual Modification:}
		Early face anonymization methods \cite{ref19, ref2, ref1, ref3, ref18} are usually conducted by visual modification,  including blurring, mosaicing, and masking. They sacrifice image quality to remove information that can be used for human perception. However, the loss of identity information and visual quality limits its application in the field of computer vision.
		
		To improve the quality of virtual faces, deep learning-based methods have been proposed. We can divide these schemes into two groups. One is identity preservation \cite{ref5, ref22, ref23, ref6, ref7, ref37, ref24}. These methods alter the appearance of faces for visual protection. However, the identity features are maintained in the protected faces, which may lead to the risk of identity theft. The other is identity modification \cite{ref21, ref8, ref9, ref20, ref25, ref11}. These anonymization methods are irreversible, as the identity of the original face is permanently lost, making it impossible to perform downstream tasks like face recognition.

		\textbf{Anonymization by Visual Preservation:}
		The virtual faces generated by such methods maintain the original visual information, i.e., not altering the appearance of the original face. Methods \cite{ref17, ref16} decouple facial attributes and modify sensitive attributes to protect privacy, ensuring that the generated face have minimal visual changes.
		Saheb et al. \cite{ref16} use semi-adversarial networks to learn the feature representation of input face images through convolutional autoencoders, ensuring that the generated faces offer maximum privacy protection with minimal distortion. 
		Mirjalili et al. \cite{ref15} utilize the method of generating adversarial perturbations to protect gender information, which can change gender information without affecting the discrimination of biometric matchers and human observers.
		Shan et al. \cite{ref13} propose a face anonymizationy method called Fawkes that can help users resist face recognition systems. This method adds imperceptible  perturbations on photos for privacy protected visual content sharing. These methods typically anonymize certain visual or identity attributes while revealing important face information.

		\textbf{Anonymization by Visual Reversibility:}
		Visually recoverable anonymization methods \cite{ref26, ref28, ref29, ref27} usually have to meet two requirements: one is anonymization and the other is de-anonymization.
		Gu et al. \cite{ref28} propose a new face identity conversion module, which can automatically perform key-based anonymization and de-anonymization of faces. Inspired by Gu et al. \cite{ref28}, Pan et al. \cite{ref29} propose a framework based on a conditional encoder and decoder, which can achieve diversity and controllability of the virtual faces according to a key and a privacy protection level. These anonymization methods realize the reversibility of identity information, allowing people with the correct key to view user data, and prevent the leakage of facial privacy to a certain extent.

		\section{Problem Statement}
		\begin{figure}[!t]
			\centering 
			\includegraphics[width=\linewidth]{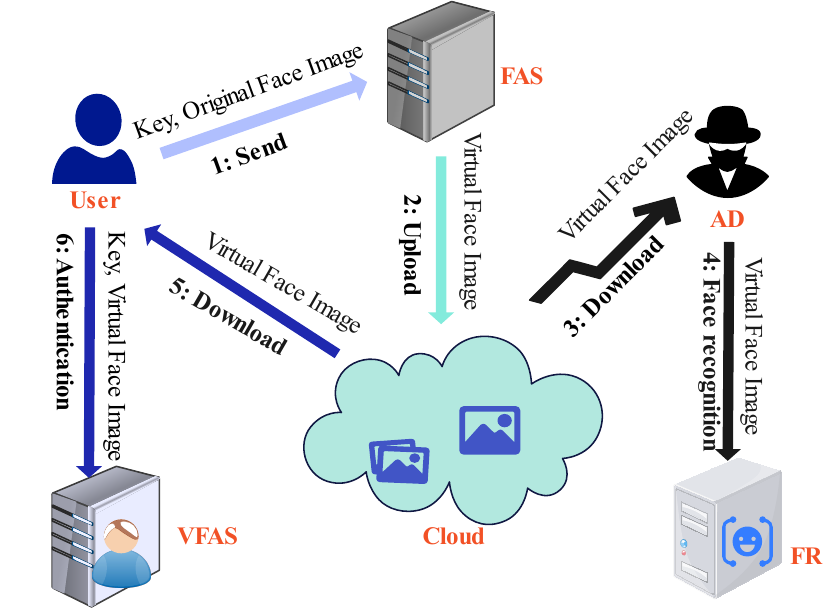}
			\caption{System Model. The participants are user (U), face anonymization server (FAS), face recognizer (FR), virtual face authentication server (VFAS) and adversary (AD).}
			\label{fig_2}    
		\end{figure}
		\subsection{System Model}
		
		The system model considered in this paper is shown in Figure \ref{fig_2}, in which the involved participants include user (U), face anonymization server (FAS), face recognizer (FR), virtual face authentication server (VFAS) and adversary (AD).

		U is responsible for securely generating the user keys. Specifically, U calls the KeyGen algorithm to generate a key and sends the user key $k$ and the original face image $x$ to FAS. For authentication, U downloads virtual face images from the cloud and sends them along with the user key to VFAS to obtain the identity of the virtual face.
		
		FAS receives the key $k $ from U and the user's face image $x$, and generates the corresponding virtual face image $x_{v}$. After the anonymization, FAS clears user information and uploads $x_{v}$ to the cloud.
		
		FR is a universal face recognition algorithm that can be accessed by anyone.
		
		VFAS is a server used for virtual face identity authentication, which receives the key $k$ sent by U and the virtual face $x_{v}$ to obtain the identity of the virtual face. If the key is correct, the identity of the virtual face is consistent with that of the corresponding original face. 
		
		AD represents all attackers attempting to obtain the original identity from the virtual faces. For AD, only the virtual faces stored in the cloud can be accessed. Therefore, AD will attempt to guess the key $k$ and obtain access control permissions of VFAS to reveal original identity of the virtual face.
		
		\subsection{The Interaction Process}
		The interaction process among different participants can be further divided into the following six stages. 
		
		\textbf{Step1:} U sends user keys and original face images to FAS.
		
		\textbf{Step2:} FAS conducts the anonymization of the face image, generates a virtual face, and uploads the virtual face to the cloud.
		
		\textbf{Step3:} The virtual images saved in the cloud are shared, and AD can easily access and download virtual face images from the cloud.
		
		\textbf{Step4:} AD can use a universal face recognition algorithm FR, but the obtained identity differs greatly from the original face.
		
		\textbf{Step5:} U downloads virtual faces from the cloud for authentication purpose.
		
		\textbf{Step6:} U sends the virtual face and user keys to VFAS to obtain the same identity as the original face for authentication.
		
		\subsection{Threat Model}
		We consider U to own the data itself and be trustworthy, and FAS as an honest entity. While VFAS is considered as an honest-but-curious entity, which follows the required protocol and performs the authentication. 
		Based on the mechanism of our model, we can determine that the identity of the original face can only be obtained if AD steals the user key and gains access to VFAS. AD may steal U's identity through the following means: 
		
		\subsubsection{Key Guessing} AD may hack the user keys through the following methods:

		\begin{itemize}
			\item {\textbf{User Information Assisted Guess.}} AD studies the user's digital footprint and attempts to guess the user's key based on such information. AD may also try common keys to guess the user keys.
			\item {\textbf{Brute Force Attack.}} AD uses robots to repeatedly use random keys until the correct key is found.
		\end{itemize}
		
		\subsubsection{Model Guessing} AD may obtain access to VFAS by guessing the model. AD may attempt to guess the model through various means, such as  using brute force attack, social engineering, phishing emails, etc., to obtain credentials or information for accessing the model.

		\section{The Proposed Method}
		
		\subsection{Overview}
		The proposed KFAAR framework consists of an HPVFG module and a KVFA module. The HPVFG module utilizes user keys to convert the latent vectors of the original faces into virtual vectors, preserving important attributes like head postures and facial expression. The KVFA module aims to verify the identity of virtual faces using the correct key without the need to restore the original face, thereby ensuring privacy.
		We use multi-task learning objectives to train HPVFG and KVFA modules, effectively achieving the dual goals of anonymization identifiability.

		\begin{figure*}[t]
			\centering 
			\includegraphics[width=\linewidth]{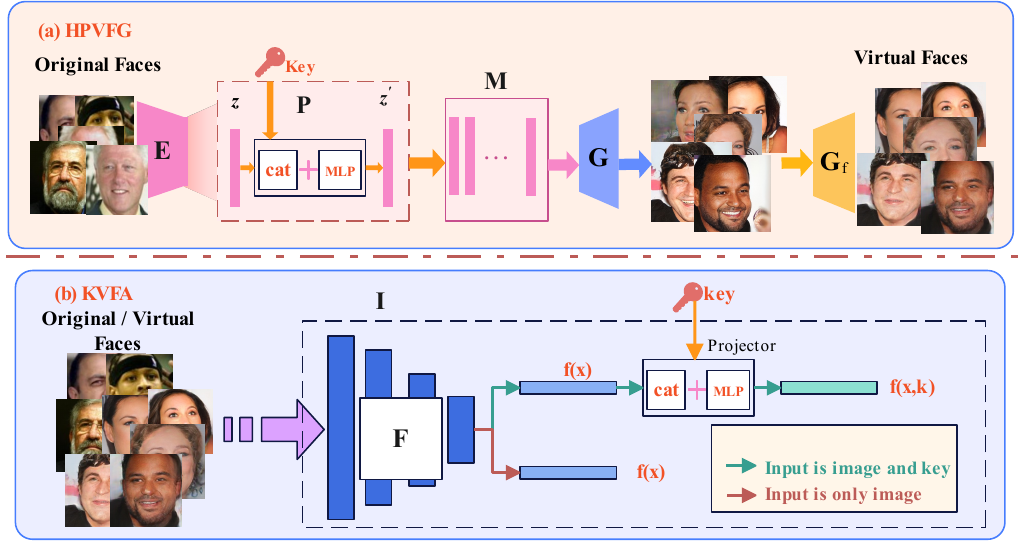}
			\caption{The proposed KFAAR framework for virtual face generation and authentication. (a) HPVFG for virtual face generation, (b) KVFA for virtual face authentication.}
			\label{fig_3}    
		\end{figure*}
		\subsection{Virtual Face Generation}
		\subsubsection{Properties of Virtual Faces}
		The goal of HPVFG is to generate a virtual face with a new appearance with the same head posture and facial expression as the original face. In addition, the virtual faces also need to be diverse, differentiated, and interactive.
		
		Given two sets of face images $\mathcal{X}$ and $\mathcal{Y}$, and the user keys $\mathcal{K}$ which control the generation of virtual faces. The generation of a virtual face can be expressed as $ G\left( x, k \right) $, and $ R\left( x \right) $ represents the feature of $ x $  extracted by a general face recognizer. Given the original faces $ x_{1} $, $ x_{2} $ $ \in $ $\mathcal{X}$, $ y $ $ \in $ $\mathcal{Y}$,  and two distinct keys  $ k_{1} $, $ k_{2} $ $ \in $ $\mathcal{K}$, our goal is to generate the virtual faces with the following properties.

		\begin{itemize}
			\item {\textit{Anonymity:} } The virtual face has a different identity from the original face, formulated as,
			
			\begin{equation}
				R\left( G\left( x_{1},k_{1} \right) \right)  \neq R\left( x_{1} \right).
			\end{equation}
			\item {\textit{Synchronism: }} If  $ x_{1} $ and  $ x_{2} $ share the same original identity, the virtual faces derived from the same key should belong to the same virtual identity, formulated as,
			
			\begin{equation}
				R\left( G\left( x_{1},k_{1} \right) \right)  = 	R\left( G\left( x_{2},k_{1} \right) \right). 
			\end{equation}
			\item {\textit{Diversity: }}The virtual identities generated by the same original face but using different keys should be different, formulated as,
			
			\begin{equation}
				R\left( G\left( x_{1},k_{1} \right) \right)  \neq 	R\left( G\left( x_{1},k_{2} \right) \right). 
			\end{equation}
			\item {\textit{Differentiation: }} For two original faces $ x $ and $ y $ which are from different identities, the generated virtual identities should be different even if they are derived from the same key, formulated as,
			
			\begin{equation}
				R\left( G\left( x,k_{1} \right) \right)  \neq 	R\left( G\left( y,k_{1} \right) \right). 
			\end{equation}
			\item {\textit{Interactivity:}} Virtual faces should be able to track the head posture and facial expression, thereby enhancing the user experience in the virtual environment. 
			
			\item {\textit{Realism:}} The authenticity of the virtual faces is essential, and the appearance of the virtual faces needs to have a high degree of realism.
			
		\end{itemize}
		
		\subsubsection{Network Architecture of HPVFG}
		As shown in Figure \ref{fig_3} (a), HPVFG consists of five parts, including: an encoder $E$, a projector $P$, a mapping network $M$, a face generator $G$ and a head posture correction module $G_{f}$.   Each part will be described in detail in this section.
		\begin{table}[t]
			\small
			\center
			
			\begin{tabular}{l}\toprule
				\textbf{Algorithm 1:} The Virtual Face Generation Process.\\ \midrule
				\textbf{Input:} The user key $k$, the original face $x$.\\
				\textbf{Onput:} The virtual face $x_{v}$.\\
				\textit{U} and  \textit{FAS}\\
				For \textit{U}\\
				\quad 1: \textit{U} generates the user key $k$.\\
				\quad 2: \textit{U} sends the user key $k$ and the original image $x$.\\
				
				For \textit{FAS}\\
				\quad 3: \textit{FAS} receives $k$ and $x$.\\
				\quad 4: Extracting features $z$ of $x$ using the Encoder \textit{E}.\\
				\quad\quad$z$ = \textit{E}($x$)\\
				\quad 5: Combining $z$ with key $k$ using Projector \textit{P} to obtain $z'$.\\
				\quad\quad $z'$ = \textit{P}($z$,$k$)\\
				\quad 6: Using the mapping network \textit{M} to map $z'$ to a multidimensional \\\quad \quad space $Z^{+}$, obtain $z^{+}$.\\
				\quad 7: Inputting $z^{+}$ into the generator \textit{G} of StyleGAN2 to obtain the \\ \quad \quad virtual face $x'$.\\
				\quad\quad $x'$ = \textit{G}($z^{+}$)\\
				
				\quad 8: Using FaceVid2Vid $G_{f}$ to perform posture correction on $x'$, \\ \quad\quad making the expression and head posture of the virtual face \\ \quad \quad consistent with the original face, and obtaining the virtual  \\ \quad  \quad face $x_{v}$.\\
				\quad\quad $x_{v}$ = $G_{f}$($x'$, $x$)\\
				
				\textit{FAS} clears user information and uploads  $x_{v}$ to the cloud.
				
				\\	\bottomrule
			\end{tabular}
		\end{table}
		
		\textbf{Encoder:} We use a pre-trained face recognizer as the encoder $E$. The face features are extracted from the original faces and represented as \( z= E \left(x\right)\).
		
		\textbf{Projector:} We propose a projector $P$ that modifies the face feature by combining the original face feature $ z $ and key $ k $.  It can be formulated as \( z'=P(z,k) \). The network consists of a concatenated operation and a multilayer perceptron (MLP). 
		
		\textbf{Mapping Network:} The main task of the mapping network is to generate style parameters, which expands the 512-dimensional latent $ z $ from the latent space $ Z $ to $ Z^{+} $. The extended latent $ z^{+} $ consists of 18 different $ z $. It can be formulated as \( z^{+}=M(z) \).

		\textbf{Generator:} This generator uses the pre-trained StyleGAN2 to reconstruct the face images. It maps style-mixed extended latent $ z^{+} $ into virtual face images. It can be formulated as \( x' = G(z^{+}) \).
		
		\textbf{Head Posture Correction Module:}
		FaceVid2Vid is a method based on conditional generative adversarial network (CGAN) \cite{ref39} for head posture replay of face images. The main idea is to map the input virtual face image and the original head posture to a shared latent variable space by learning the relationship between the face image and the corresponding head posture, thereby achieving the transformation of the head head posture. The FaceVid2Vid model can replay head postures by learning the relationship between face images and corresponding head posture. This model can be applied to fields like virtual reality, providing users with a more realistic experience.
		
		In order to make the virtual faces have the posture and expression of the original faces, we use a pre-trained face reconstruction model FaceVid2Vid to obtain the virtual faces $x_{v}$. The facial reconstruction process can be described as: \(x_{v}=G_{f} (x', x) \), where $G_{f}$ represents the FaceVid2Vid model.
		
		The encoder $E$ takes a face image tensor as input, the projector $P$ combines $k$ with the face feature and projects it to the latent space of StyleGAN2. $G$ generates a realistic virtual face with virtual identity. The virtual face learns the head posture and facial expression features of the original face through $x_{v}=G_{f}$ to generate a virtual face that maintains the head posture and face expression (see Algorithm 1 for the generation process).
		
		\subsubsection{Training of HPVFG}
		We use a multi-task learning approach to train HPVFG, as shown in Figure 4 (a), which can achieve the first four properties of virtual faces. In the training, the projector $P$ is trainable, while others are frozen.
		We use the cosine embedding loss $ L_{cos} $ to measure the similarity between the face features. It can be expressed as:
		\begin{figure}[!t]
			\centering 
			\includegraphics[width=\linewidth]{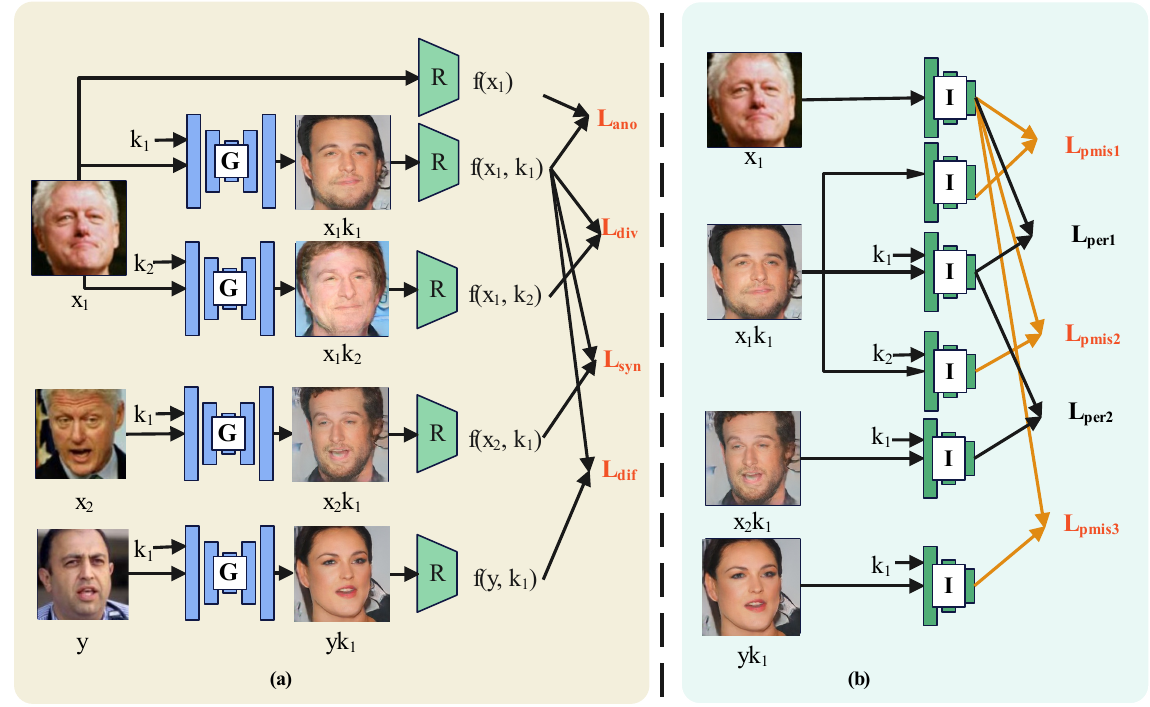}
			\caption{Illustration of our training strategy. (a) Virtual Face Generation, (b)  virtual Face Authentication.}
			\label{fig_4}    
		\end{figure}
		
		\begin{equation}
			\textit{L}_{\cos}\left(f_{1}, f_{2},l \right)  = \left\{\begin{matrix}
				\left(1-\cos\left(f_{1}, f_{2} \right) \right) ,&l=1, \\
				\max\left( m,\cos\left(f_{1}, f_{2}\right) \right),&l=-1,
			\end{matrix}\right.
		\end{equation}
		where $ m $ represents a hyperparameter, $ l=1 $ means that $f_{1}$ and $f_{2}$ come from faces with the same identity, $ l=-1 $ means that $f_{1}$ and $f_{2}$ are from different faces. We optimize $ \textit{L}_{cos} $ to maximize/minimize the difference between features.
		
		\begin{itemize}
			\item {\textit{Anonymity Loss. }}
			We hope that the original face and the virtual face are not belong to the same identity, and the ID-distance between them is increased by minimization $ \textit{L}_{ano} $, which can be expressed as:
			\begin{equation}
				\textit{L}_{ano} = \textit{L}_{\cos}\left(R\left( G \left(x_{1},k_{1}\right)  \right),R\left(x_{1} \right),-1  \right). 
			\end{equation}
			
			\item {\textit{Synchronism Loss.}} Given two different face images $ x_{1},x_{2}\in \mathcal{X}  $ corresponding to the same identity, the corresponding virtual faces generated, with the same key, have the same identity. We propose $ \textit{L}_{syn} $ loss to represent reducing the identity distance between virtual faces:
			\begin{equation}
				\textit{L}_{syn} = \textit{L}_{\cos}\left(R\left( G \left(x_{1},k_{1}\right)  \right), R\left( G \left(x_{2},k_{1}\right)  \right), 1  \right).   
			\end{equation}
			
			\item {\textit{Diversity Loss.}} To ensure the diversity of virtual faces, we attempt to ensure that the virtual faces generated from the same face image controlled by different keys have different identities. We propose $ \textit{L}_{div} $, given by:
			\begin{equation}
				\textit{L}_{div} = \textit{L}_{\cos}\left(R\left( G \left(x_{1},k_{1}\right)  \right), R\left(G \left(x_{1},k_{2}\right)  \right), -1  \right).   
			\end{equation}
			
			\item {\textit{Differentiation Loss.}} To meet the differentiation property, the virtual faces generated from different original faces should belong to different identities. Given two face images $ x \in \mathcal{X}$ and $ y \in \mathcal{Y}$, to generate virtual faces with different identities using the same key, it is necessary to reduce the decisive role of the key on the identity. We define $ \textit{L}_{dif} $ and expand the differences between virtual faces by minimizing this loss, given by:
			\begin{equation}
				\textit{L}_{dif} = \textit{L}_{\cos}\left(R\left( G \left(x,k_{1}\right)  \right), R\left( G \left(y,k_{1}\right)  \right), -1  \right). 	 
			\end{equation}
			
			\item {\textit{Total Loss.}} Our overall optimization goal is the weighted average of the above losses, which is given below. 
			\begin{equation}
				\textit{L}_{tot} = \lambda_{ano} \textit{L}_{ano}+\lambda_{syn}  \textit{L}_{syn} + \lambda_{div}  \textit{L}_{div} +\lambda_{dif}  \textit{L}_{dif}, 
			\end{equation}
			where $ \lambda_{ano} $, $ \lambda_{syn} $, $ \lambda_{div} $ and $ \lambda_{dif} $ are the weights of different losses.
		\end{itemize}

		\subsection{Authentication}
		\subsubsection{Properties of KVFA}
		To ensure that the virtual faces are traceable, we train the KVFA network to ensure that the virtual face can be authenticated by the key. We use $ I\left( x \right) $ to denote the identity of $ x $ extracted by this network. It has the following properties:

		\textbf{Prevent Misidentification: }
		\begin{itemize}
			\item{The original face and the virtual face belong to different identities.
				\begin{equation}
					I\left(x\right)  \neq 	I\left( G\left( x,k_{1} \right) \right). 
			\end{equation}}
			
			\item{When using the wrong key for authentication, the identities of the original face and the virtual face are different.
				\begin{equation}
					I\left(x\right)  \neq 	I\left( G\left( x,k_{1} \right), k_{2}\right). 
			\end{equation}}
			
			\item{When using the correct secret key for authentication, different faces have different identities.
				\begin{equation}
					I\left( G\left( x,k_{1} \right),k_{1} \right) \neq 	I\left( G\left( y,k_{1} \right), k_{1}\right). 
			\end{equation}}
		\end{itemize}

		\textbf{Prevent False Rejection:}
		\begin{itemize}
			\item{When using the correct key for authentication, the identities of the original face and the virtual face are consistent.
				\begin{equation}
					I\left(x\right)  = 	I\left( G\left( x,k \right),k \right). 
			\end{equation}}
			
			\item{When using the correct key for authentication, two different virtual faces from the same original identity belong to the same identity.
				\begin{equation}
					I\left( G\left( x_{1},k_{1} \right),k_{1} \right) =	I\left( G\left( x_{1},k_{2} \right), k_{2}\right),
				\end{equation}
				and
				\begin{equation}
					I\left( G\left( x_{1},k_{1} \right),k_{1} \right) =	I\left( G\left( x_{2},k_{2} \right), k_{2}\right). 
			\end{equation}}
		\end{itemize}
		
		\subsubsection{Network Architecture of KVFA} KVFA (as shown in Figure 3 (b)) consists of two parts, including a feature extractor (F) and a projector (P). When KVFA inputs an image, the output of the model is the feature representation of the image. When the input of HPVFG is an image and a key, the output is a projection of the feature representation of the image combined with the key.
		
		\begin{itemize}
			\item {\textbf{Feature Extractor:}} We use an end-to-end network as a feature extractor, which maps the input image directly to the feature representation. This network consists of multiple layers, including the input layer, hidden layer, and output layer, and we set the number of hidden layers as 4. Feature representations from input images can be described as \( z_{v} = F (x_{v}) \).
			
			\item {\textbf{Projector:}} We propose a projector \textit{P} that modifies the face feature by combining the face feature $ z_{v} $ and key $ k $.  It can be formulated as \(z_{v}'=P(z_{v},k) \). The network consists of an MLP. 
		\end{itemize}
		
		Using KVFA, we can achieve traceability of virtual faces using the correct key without restoring the original faces, the authentication process is summarized in Algorithm 2.

		\begin{table}[t]
			\small
			\center
			\begin{tabular}{l}\toprule
				\textbf{Algorithm 2:}  The Face Authentication Process. \\ \midrule
				\textbf{Input:} The Original face $x$, the user key $k$ and the virtual\\ face $x_{v}$.\\
				\textbf{Onput:} The result of identity authentication.\\
				\textit{U} and \textit{VFAS}\\
				\textbf{Condition 1:}\\
				For  \textit{U}\\
				\quad 1: \textit{U} sends the original image $x$ and virtual image $x_{v}$ to \textit{VFAS}.\\
				
				For  \textit{VFAS}\\
				\quad 2: \textit{VFAS} only receives  $x$ or $x_{v}$.\\
				\quad 3: Extracting the features $f_{x}$ ($f_{v}$) of $x$ ($x_{v}$) using the feature \\\quad extractor \textit{F}.\\
				\quad\quad  $f_{x}$ = $z_{x}$ = \textit{F}($x$), $f_{v}$ = $z_{v}$ = \textit{F}($x_{v}$). \\
				For \textit{U}\\
				\quad 4: Comparing similarity between $f_{x}$ and $f_{v}$ and output identity \\ \quad \quad authentication results\\ 
				\textbf{Condition 2:}\\
				For \textit{U}\\
				\quad 1: \textit{U} sends the original image $x$, the user key $k$ and the virtual \\ \quad \quad image $x_{v}$ to \textit{VFAS}.\\
				
				For \textit{VFAS}\\
				\quad 2: \textit{VFAS} receives  $x$, $k$ and $x_{v}$.\\
				\quad 3: Extracting features $z_{v}$ of $x_{v}$ using the Feature extractor \textit{F}.\\
				\quad\quad  $f_{x}$ = $z_{x}$ =\textit{ F}($x$),  $z_{v}$ =\textit{ F}($x_{v}$).\\
				\quad 4: Combining $z_{v}$ with the key $k$ to obtain $z_{v}'$ and through MLP  \\ \quad \quad to get the virtual facial representation $f$. \\
				\quad\quad$z_{v}'$ = Cat($z_{v}$, $k$);\\
				\quad\quad$f$ = MLP($z_{v}'$).\\
				For \textit{U}\\
				\quad 5: Comparing the similarity between $f_{x}$ and $f_{v}$ and output the \\ \quad \quad identity authentication results
				\\	\bottomrule
			\end{tabular}
		\end{table}
		\subsubsection{Training of KVFA}
		We use a multi-task learning approach to train KVFA separately, as shown in Figure \ref{fig_4} (b).
		In KVFA, our goal is to reveal the original identity of the virtual face by the correct key, and the appearance  of the original face is not exposed during the authentication. Since virtual faces are anonymous, universal face recognizers are not applicable. We train a specialized key-conditioned face recognizer $ I $ to extract features of virtual faces. When the input key is correct, the features of the virtual face and the original face are consistent. We use $ \textit{L}_{\cos} $ (5) to measure the similarity of face features.
		
		\textbf{Prevent Misidentification: }First of all, the model should have the ability to prevent misrecognition, that is, to prevent two different faces from being recognized as the same identity. We design set of losses to meet the needs, the details of which are given below for different cases.
		\begin{itemize}
			\item{\textbf{Case 1:}} This model can distinguish the original face and the virtual face, i.e., their features of are different, the corresponding loss can be expressed as:
			
			\begin{equation}
				\textit{L}_{pmis1} = \textit{L}_{\cos}\left(I\left( G \left(x_{1},k_{1}\right)  \right),I \left(x_{1} \right),-1  \right). 
			\end{equation}
			
			\item{\textbf{Case 2:}} The authentication of virtual faces is controlled by the correct key. When an incorrect key is used as input, their features should be different using $I$. The corresponding loss is defined as follows:
			\begin{equation}
				\textit{L}_{pmis2} = \textit{L}_{\cos}\left(I \left( G \left(x_{1},k_{1}\right) ,k_{2} \right),I \left(x_{1} \right),-1  \right). 
			\end{equation}
			
			\item{\textbf{Case 3:}} For virtual faces from different original identities, even if the correct secret key is used for authentication, their features should be different using $I$. The corresponding loss is defined as follows:
			\begin{equation}
				\textit{L}_{pmis3}=\textit{L}_{\cos}\left(I\left( G \left(x,k_{1}\right) ,k_{1} \right),I \left( G \left(y,k_{1}\right) ,k_{1} \right),-1  \right). 
			\end{equation}
			\item{\textbf{Total Loss to Prevent Misidentification.}} The total loss to prevent misidentification can be described as:
			\begin{equation}
				\textit{L}_{tot1} = \lambda_{pmis1}  \textit{L}_{pmis1} +\lambda_{pmis2}  \textit{L}_{pmis2} + \lambda_{pmis3}  \textit{L}_{pmis3},
			\end{equation}
			where $\lambda_{pmis1}$, $\lambda_{pmis2}$, and $\lambda_{pmis3}$ are the weights for different losses. 	\end{itemize}
		
		\textbf{Prevent False Rejection: } This model can prevent faces of the same identity from being falsely rejected. We design a set of different losses for different cases, which is detailed below.
		\begin{itemize}
			
			\item{\textbf{Case 1:}} When the correct key is used, the identities of the virtual face and the original face are consistent. We minimize the distance between their features by optimizing $ \textit{L}_{per1} $, which can be described as:
			\begin{equation}
				\textit{L}_{per1} = \textit{L}_{\cos}\left(I\left( G \left(x_{1}, k_{1}\right), k_{1} \right),I \left(x_{1} \right),1  \right).
			\end{equation}
			
			\item{\textbf{Case 2:}} Different virtual faces generated from the same identity should have the same identity when the correct secret key is presented. We minimize the distance between the two virtual faces' features by optimizing $ \textit{L}_{per2} $, which can be described as:
			\begin{equation}
				\textit{L}_{per2} = \textit{L}_{\cos}\left(I\left( G \left(x_{1},k_{1}\right) ,k_{1} \right),I\left(G \left(x_{2},k_{1}\right) ,k_{1} \right),1  \right).
			\end{equation}
			\item {\textbf{Total Loss to Prevent False Rejection: }} The total loss to prevent false rejection can be described as:
			\begin{equation}
				\textit{L}_{tot2} = \lambda_{per1}  \textit{L}_{per1}+\lambda_{per2}  \textit{L}_{per2},
			\end{equation} 
			
			where$ \lambda_{per1}$ and $\lambda_{per2}$ are the weight for different losses.	\end{itemize}
		
		\textbf{Overall Objective.} The overall  objective for KVFA is formulated as follows:
		\begin{equation}
			\textit{L}_{tot} = \textit{L}_{tot1}+ \textit{L}_{tot2}.
		\end{equation}
		
		\section{Experimental Results}
		In this section, we conduct a series of experiments to evaluate the effectiveness of the proposed framework.
		
		\subsection{Experimental Setting} 
		$\bullet$ \textit{Dataset:}
		We conduct experiments on the following two public datasets: 
		(1) LFW \cite{ref30}, which contains a total of more than 13,000 face images from 13,233 individuals, of which 1,680 have more than two images. (2) CelebA \cite{ref31}, which contains 202,599 face images of 10,177 celebrities. 
		
		$\bullet$ \textit{Pre-trained Models:}
		In the training phase, we use FaceNet Inception-ResNet-v1 \cite{ref32} which is trained on VGGFace2 \cite{ref33} for both encoder $ E $ and recognizer $ R $. 
		We train the official release of StyleGAN2 on the LFW and CelebA datasets as our generator $ G $.
		We use the pre-trained FaceVid2Vid as the head posture correction module   $G_{f}$. 
		During the testing and evaluation phase, we choose the pre-trained ArcFace \cite{ref34} and Sphereface \cite{ref35} as face recognizers $ R $.
		
		$\bullet$ \textit{Parameter Setting:} The projector is trained for 10 epochs by the Adam optimizer with $ \beta_{1} = 0.9 $ and $ \beta_{2} = 0.999 $.  Learning rate is set to $ 1 \times 10^{-4} $, and we set $\lambda_{ano}=0.4$, and $\lambda_{syn}= \lambda_{div}= \lambda_{dif}=1$. KVFA is also used by the Adam optimizer with $\beta_{1}=0.9$ and $\beta_{2}=0.999$, where the batch size is set as 2, the learning rate is set as $ 1 \times 10^{-4} $ and $\lambda_{pmis1}=\lambda_{pmis2}=\lambda_{pmis3}=\lambda_{per1}=\lambda_{per2}=1$. 
		The training on LFW takes four hours on a single NVIDIA GTX 3090 GPU.

		$\bullet$ \textit{State of the Art:}
		We compare our anonymization framework HPVFG with three state-of-the-art (SOTA) methods, namely CIAGAN \cite{ref6}, IVFG \cite{ref12}, and VFGM \cite{ref37}.
		
		$\bullet$ \textit{Evaluation Metrics:} To evaluate the anonymity of the virtual face, we use "anonymization" to measure the unsuccessful matching between the original face and the virtual faces. We use equal error rate (EER) and area under curve (AUC) to evaluate the synchronism of virtual faces. AUC measures the area under the
		receiver operating characteristic (ROC) curve. Larger AUC and smaller EER indicate higher the accuracy of the face recognition system. We use the objective image quality assessment metrics, i.e., Frech’et inception distance (FID), to evaluate the visual quality of the virtual faces. In addition, to evaluate the performance of head posture and facial expression preservation, we use existing head posture estimation techniques to calculate the Euler angles (Yaw, Pitch, and Roll) within the human body.
		
		\begin{figure*}[t]
			\centering 
			\includegraphics[width=\linewidth]{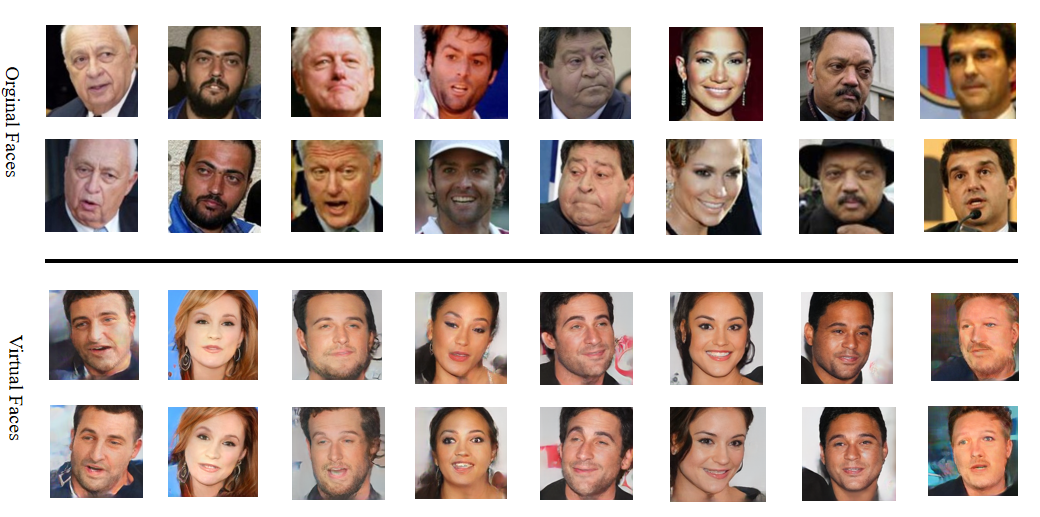}
			\caption{Examples of the virtual faces. The first two rows are the original face images, while the third and fourth rows are the virtual face images of the first and second rows, respectively.}
			\label{fig_5}    
		\end{figure*}
		
		\begin{table}[t]
			\caption{Anonymity and diversity of virtual face images generated by different methods.}
			\center
			\label{tab_2}
			\begin{tabular}{ccccc}\toprule
				\multicolumn{1}{c}{\multirow{2}{*}{Methods}} & \multicolumn{2}{c}{Anonymity  $ \uparrow $} &\multicolumn{2}{c}{ Diversity$\uparrow$}\\ 
				\cmidrule(r){2-3} \cmidrule(r){4-5}
				& LFW & CelebA & LFW & CelebA \\\midrule
				CIAGAN \cite{ref6}	&0.674  & 0.0628&0.000   &0.000\\
				IVFG \cite{ref12}	&\textbf{0.988} &0.889 & 0.750   &\textbf{0.787}\\
				VFGM \cite{ref37}	& 0.524 & 0.561 & 0.550 &0.532\\
				Ours &0.962 &\textbf{0.922} &\textbf{0.783}  &0.728	
				\\	\bottomrule
			\end{tabular}
		\end{table}
		
		\begin{figure*}[!t]
			\centering 
			\includegraphics[width=\linewidth]{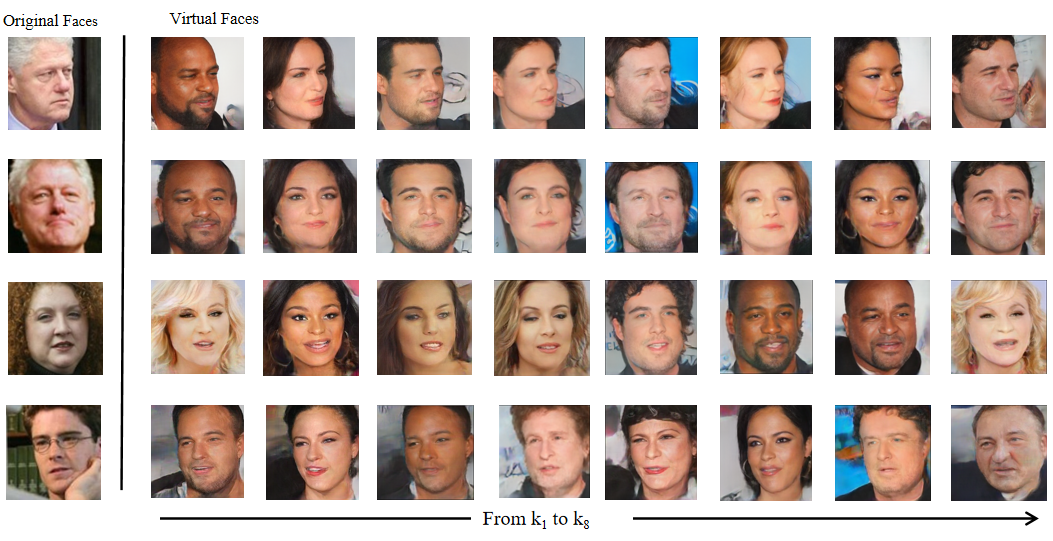}
			\caption{Diverse anonymization results. The leftmost column represents the original faces, and rest columns illustrate different virtual faces using different keys. The virtual faces in each column share the same key.}
			\label{fig_6}    
		\end{figure*}
		
		\begin{table}[t]
			\caption{Synchronism and detection rate of the virtual faces generated using different methods.}
			\center
			\label{tab:3}
			\begin{tabular}{ccccccc}\toprule
				\multicolumn{1}{c}{\multirow{2}{*}{Methods}} & \multicolumn{2}{c}{AUC $\uparrow$}  &\multicolumn{2}{c}{ EER $ \downarrow $ }&\multicolumn{2}{c}{  Detection Rate $\uparrow$ } \\ 
				\cmidrule(r){2-3} \cmidrule(r){4-5} \cmidrule(r){6-7}			
				& LFW  &  CelebA	& LFW  &  CelebA 	& LFW  &  CelebA \\\midrule
				CIAGAN \cite{ref6}&-&-&-&-&0.986&0.998\\
				IVFG \cite{ref12}	&0.929 & 0.933 & 0.103&0.122 &1.0         &1.0 \\
				VFGM \cite{ref37}	&0.889 &0.920& 0.182 &0.139& 1.0         &1.0\\
				Ours &\textbf{0.973} &\textbf{0.992}&\textbf{0.092} &\textbf{0.089}&\textbf{1.0}& \textbf{1.0}
				\\	\bottomrule
			\end{tabular}
		\end{table}

		\subsection{Evaluation of Virtual Faces}
		We evaluate the performance of our proposed framework in terms of anonymity, diversity, synchronism, detection rate, interactivity, and visual quality, and compare these performances with SOTA methods.
		
		\subsubsection{Anonymity and Diversity} As shown in Figure \ref{fig_5}, we give examples of the original faces and the corresponding virtual faces, where the first two rows are original faces from the same identity, and the last two rows correspond to virtual faces. It can be seen that there are significant visual differences between the original face and the virtual face, and the virtual face maintains the head posture and facial expression of the original face. We quantify the anonymity of the virtual face by calculating the mismatch rate between the virtual face and the original face. When the distance between features of the original face and the virtual face is less than the threshold of the face recognizer, we regard them to be from different identities. Table \ref{tab_2} summarizes the performance of different methods on the LFW and CelebA datasets. Our method achieve a high anonymization  of around 0.98, which significantly outperforms CIAGAN \cite{ref6} and VFGM \cite{ref37}, and is similar to IVFG \cite{ref12}.
		
		In addition, we explore the effects of using different keys to generate virtual face images. We calculate the unsuccessful matching rate between virtual face images generated using different keys. The results are shown in Table \ref{tab_2}, the diversity is superior to SOTA on the LFW dataset, and is only 0.059 lower than IVFG \cite{ref12} on the CelebA dataset. Figure \ref{fig_6} shows eight virtual face images generated using eight different keys from four original face images, respectively. It can be seen that under different keys, virtual faces have significant differences in appearance. 
		
		\subsubsection{Synchronism and Detection Rate}
		One of the goals of our method is to generate virtual face images that have high utility, including high synchronism (the original faces of the same identity are controlled by the same key to generate the virtual face images with the same virtual identity) and high detection rate. We use randomly generated 8-bit keys and the LFW testing set as inputs for HPVFG to generate the corresponding virtual face images. We can see from Figure \ref{fig_5} that different original faces of the same identity are able to generate virtual faces with the same virtual identity by using the same key. At the same time, the virtual faces maintain the head posture and facial expression of the original face.

		\begin{table*}[t]
			\caption{The ability of different methods in terms of preservation of head posture and facial expression.}
			\center
			\label{tab:4}
			\begin{tabular}{ccccccccc}\toprule
				\multicolumn{1}{c}{\multirow{2}{*}{Methods}} & \multicolumn{2}{c}{Yaw $\downarrow$}  &\multicolumn{2}{c}{ Pitch $ \downarrow $ }&\multicolumn{2}{c}{ Roll $\downarrow$ }&\multicolumn{2}{c}{ Emotion $\uparrow$ } \\ 
				\cmidrule(lr){2-3} \cmidrule(lr){4-5} \cmidrule(lr){6-7}		\cmidrule(lr){8-9}	
				& LFW  &  CelebA 	& LFW  &  CelebA 	& LFW  &  CelebA & LFW  &  CelebA\\\midrule
				
				CIAGAN \cite{ref6}&3.913&3.152&4.387&4.424&3.037&3.291&0.584&0.622\\
				
				VFGM \cite{ref37}&3.557&3.623&\textbf{2.519}&2.333&4.011&3.997&0.726&0.689\\
				IVFG \cite{ref12} &25.663&28.537&19.334&18.261&17.991&17.309&0.433&0.412\\
				Ours &\textbf{2.018} &\textbf{1.992}&3.102 &\textbf{2.119}&\textbf{1.998}& \textbf{2.006}&\textbf{0.805}& \textbf{0.833}\\	\bottomrule
			\end{tabular}
		\end{table*}
		To demonstrate the superiority of our proposed method, we also conducted a quantitative analysis that compares the performance with SOTA on the LFW and CelebA datasets,
		we choose AUC and EER as the criteria for measuring synchronism, and use the MTCNN model to detect the virtual faces for measuring the detection rate. The results of synchronism and detection rate are shown in Table \ref{tab:3}. It can be seen that our proposed method outperforms SOTA in both aspects.		
		
		\subsubsection{Interactivity} Interactivity is one of the important properties of our virtual faces, which requires the virtual face to have the same head posture and facial expression as the original face. As shown in Figures \ref{fig_5} and \ref{fig_6}, the virtual face maintains the head posture and facial expression of the original face, demonstrating the visual interactivity of our method.
		
		In addition, we utilize the API of Face++ to measure the preservation of head posture and facial expression. Specifically, we detect the similarity between head posture angles (Yaw, Pitch, and Roll) and facial expression. As shown in Table \ref{tab:4}, the offset angle between our virtual face and the original faces in each direction is the lowest among different schemes. The similarity of facial expression can reach 0.8, which is higher than that of SOTA.
		
		\subsubsection{Visual Quality} We use FID to quantitatively evaluate the visual quality of the generated images. Table \ref{tab:5} shows FID of the virtual face images generated by different methods. It can be seen that our method achieves good image quality. It can also be seen from Figure \ref{fig_5} and Figure \ref{fig_6} that our virtual face images have high perceptual quality. 
		
		\subsubsection{Summary} Compared with SOTA, our virtual faces achieves better performance in anonymity, diversity, synchronism and detection rate, as well as the ability in preservation of head posture and facial expression. Although IVFG \cite{ref12} achieves higher visual quality, it still has the following limitations. (1) Its virtual faces have random head postures and expressions. (2) The recognition performance of IVFG \cite{ref12} is only reflected in determining whether the virtual faces come from the same original identity, and cannot be used for source tracking tasks. 
		CIAGAN \cite{ref6} has made some progress in preserving head posture, but its virtual faces are often suffer from poor visual quality and  the original identity is permanently lost, making it impossible to perform downstream recognition tasks. VFGM \cite{ref37} visually protects the original face while maintaining the head posture. It satisfies the need for interactivity. However, the lack of feature-level protection results in the original identity being arbitrarily and illegally used and recognized without authorization. Our HPVFG generates the virtual faces with high visual quality while maintaining head posture and facial expression. In addition, our proposed KVFA can achieve authorized recognition of the original identity, which prevents the virtual faces from illegal access and identity abuse. This breakthrough protects the original face in both visual and machine perception, which mitigates the contradictory between privacy and face authentication.
		
		\begin{table}[t]
			\caption{ FID of the virtual face images generated using different methods.}
			\center
			\label{tab:5}
			\begin{tabular}{ccc}\toprule
				Methods & LFW          & CelebA \\ \midrule
				CIAGAN \cite{ref6} &7.64 &6.90 \\
				VFGM \cite{ref37}	&6.99          &6.91\\
				IVFG \cite{ref12}	&\textbf{6.19} &\textbf{6.78} \\
				
				Ours    & 7.29         &6.82
				\\	\bottomrule
			\end{tabular}
		\end{table}
		
		\subsection{Identity Authentication}
		In this section, we evaluate the recognition accuracy and identity authentication ability of the KVFA module separately. We use cosine similarity as a measure of the similarity between anonymous face features and original face features. We set the threshold as 0.7, which means when the similarity is greater than 0.7, the original and the virtual faces belong to the same identity.
		
		\subsubsection{Recognition Accuracy of KVFA} 
		We use a series of evaluation indicators: correct recognition rate (CRR), false acceptance rate (FAR), and AUC to evaluate the recognition accuracy of KVFA. We conduct tests on the LFW and CelebA datasets respectively, with the inputs of KVFA being the original faces, or the virtual faces with correct keys. The experimental results are shown in Table \ref{tab:6}. It can be observed that our KVFA module performs well on the two datasets, which demonstrates its ability in tracing the original identities of the virtual faces.
		
		\begin{table}[t]
			\caption{The recognition accuracy of KVFA.}
			\center
			\label{tab:6}
			\begin{tabular}{ccccccccc}\toprule
				Dataset & CRR$ \uparrow $ & FAR$ \downarrow $ & AUC $ \uparrow $\\ \midrule
				LFW & 0.927 & 0.078 & 0.984\\
				CelebA & 0.956 & 0.064 & 0.989 \\ 
				\bottomrule
			\end{tabular}
		\end{table}
		\subsubsection{Different Authentication Scenarios}
		During the authentication, the following scenarios may occur:
		\begin{table}[t]
			\caption{The feature similarity between the original face and virtual face in different authentication scenarios. }
			\center
			\label{tab:7}
			\begin{tabular}{ccc}\toprule
				&LFW& CelebA \\ \midrule
				Scenario  1	&0.498&0.516 \\
				Scenario  2	&0.326&0.441 \\
				Scenario  3	&0.522&0.398 \\
				Scenario  4	&\textbf{0.852} &\textbf{0.799} \\
				\bottomrule
			\end{tabular}
		\end{table}
		
		\begin{table}[t]
			\caption{Results of the "In-the-wild" experiment.}
			\centering
			\label{tab:wild}
			\begin{tabular}{ccccccc}\toprule
				\multicolumn{1}{c}{\multirow{2}{*}{Datasets}} & \multicolumn{4}{c}{ \makecell{Performance \\  of the virtual faces} }  &\multicolumn{2}{c}{\makecell{Recognition\\ accuracy of KVFA }  } \\ 
				\cmidrule(lr){2-5}  \cmidrule(lr){6-7}
				&EER& Anonymity& Diversity&FID&CRR&FAR\\
				\midrule
				LFW  & 0.092& 0.962&0.783&7.29 &0.927&0.078\\
				FFHQ & 0.089& 0.931&0.707&6.36&0.887&0.081	\\
				\bottomrule
			\end{tabular}
		\end{table}
		
		\begin{table}[t]
			\caption{ Ablation studies for each loss of HPVFG.}
			\center
			\label{tab:8}
			\begin{tabular}{ccccccc}\toprule
				Methods&AUC& EER &Detection&Anonymity&Diversity&FID\\ \midrule
				w/o $\textit{L}_{ano} $&0.914 &0.091 &1.0 &0.908 &0.779 &6.86\\
				w/o $\textit{L}_{syn}$&0.897 &0.104 & 1.0&0.974& 0.787&7.73\\
				w/o $ \textit{L}_{div}$&0.942 & 0.091&1.0 & 0.962& 0&7.91\\
				w/o $\textit{L}_{dif}$& 0.917& 0.095&1.0 & 0.953&0.790 &8.04\\
				\midrule
				Ours&0.973 &0.092 &1.0 &0.962 &0.783 &7.29\\
				\bottomrule
			\end{tabular}
		\end{table}
		\begin{table}[t]
			\caption{ Ablation studies for each loss of KVFA.}
			\center
			\label{tab:9}
			\begin{tabular}{cccc}\toprule
				Methods&CRR& FAR &AUC\\ \midrule
				w/o $\textit{L}_{tot1} $&0.756 &0.298 &0.833 \\
				w/o $\textit{L}_{tot2}$&0.636 &0.270 & 0.744\\
				\midrule
				Ours&0.945 &0.068 &0.987\\
				\bottomrule
			\end{tabular}
		\end{table}
		\textbf{Scenario 1: }The adversary has the correct key, but does not have access to KVFA. They input the correct key and virtual face into HPVHG to recover the real face. Then they use a universal face recognizer to obtain face feature and attempt to match it with the feature of original face.
		
		\textbf{Scenario 2: } The adversary bypasses the key and directly input the virtual face into the KVFA module to obtain the face feature, and attempts to match it with the original face.  
		
		\textbf{Scenario 3: }The adversary inputs the wrong key and the virtual face into KVFA, to obtain the face feature, and attempts to match it with the original face. 
		
		\textbf{Scenario 4: }The adversary/user holds the correct key, inputs the virtual face and key into KVFA to obtain the feature and matches it with the original face.
		
		Table \ref{tab:7} gives the similarity between the virtual face and the original face under different authentication scenarios. It can be observed that, when adversary does not have access to KVFA, the cosine similarity between features of the recovered face and the original face is low, indicating a failed authentication. When adversary attempts to bypass the key or conduct the authentication with the wrong key, the cosine similarity between the features of the original face and the virtual face is lower in both LFW and CelebA, indicating a failure of authentication. Therefore, our method can resist illegal authentication by bypassing keys or using incorrect keys. When using the correct key for authentication, the cosine similarity between the features of the original and virtual face is over 0.8, meaning that successful authentication can be carried out by using the correct keys.


		\begin{figure}[!t]
			\centering 
			\includegraphics[width=\linewidth]{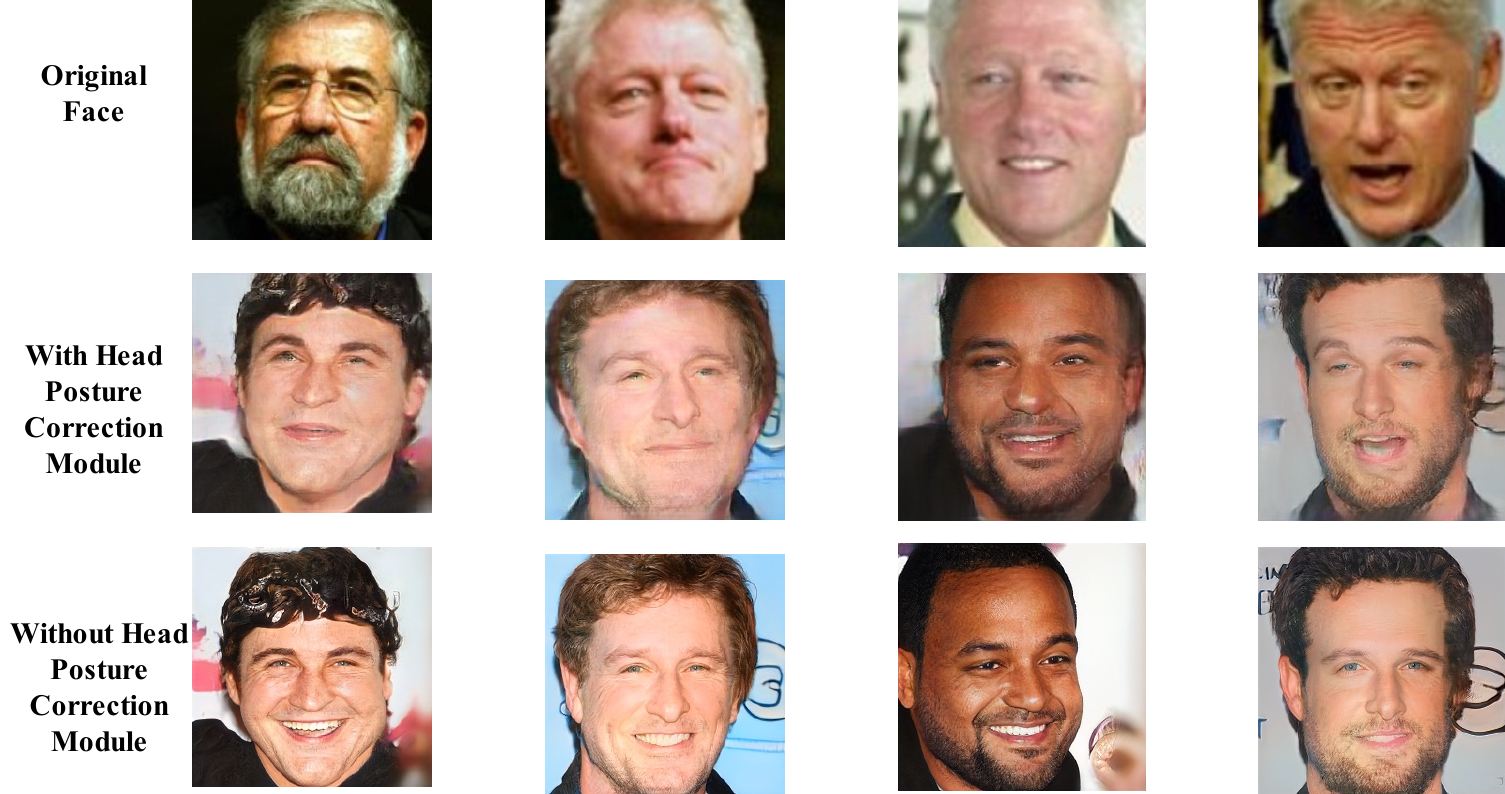}
			\caption{Ablation study for the head posture correction module.}
			\label{fig_7}    
		\end{figure}
		
		\begin{figure}[!t]
			\centering 
			\includegraphics[width=\linewidth]{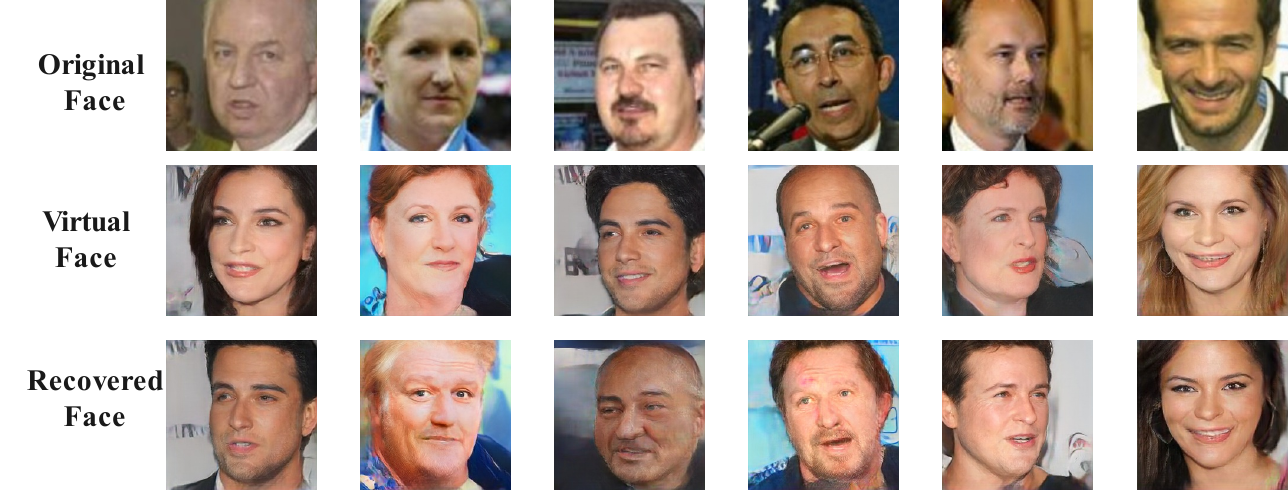}
			\caption{Examples of the recovered faces by using HPVFG.}
			\label{fig_8}    
		\end{figure}
		
		\subsection{"In-the-wild" Experiment}
		To validate the generalization ability of our model, we further select the FFHQ dataset for evaluation. FFHQ is a high-quality face image dataset that contains 70,000 high-resolution face images with a resolution of 1024$\times$1024. We randomly select 100 images from the FFHQ dataset as the validation set.
		We resize the face images to the same size as the inputs of our model trained on LFW. After resizing, we input the 100 face images into our HPVFG (trained on LFW) to obtain the virtual faces. For each original face image, we assign eight different keys to generate eight virtual faces. As such, we have 800 virtual faces in total. Table \ref{tab:wild} gives the performance of the virtual faces as well as the recognition accuracy of KVFA (trained on LFW). It can be seen that we can obtain lower FID on the FFHQ dataset thanks to the high quality face images in FFHQ. For the other performance indicators, however, the results on the FFHQ are slightly lower than those on the LFW dataset. This may be attributed to the more diverse face features and more complex backgrounds in the FFHQ dataset. But the overall performance is still higher than SOTA, which demonstrates the good generalization ability of our method among different datasets.

		\subsection{Ablation Studies}
		\subsubsection{Head Posture Correction Module} Here, we demonstrate the importance of the posture correction module in maintaining head posture and facial expression. Without this module, the identity, posture, and the expression of the virtual face generated by the generator are highly controlled by the key. As shown in Figure 7 that the head posture and facial expression of the virtual face and the original face are inconsistent without the attitude correction module. That is, under the same key, the virtual faces generated from different face images of the same identity not only belong to the same virtual identity but may also have the same head posture and facial expression. This causes the virtual face to lose the diversity of head posture and expression, which cannot meet the interactivity of the metaverse.

		\subsubsection{Each Loss of HPVFG}
		We remove each loss component from the overall training objective of HPVFG and summarize the corresponding performance in Table \ref{tab:8}. It can be seen that, by using all the loss functions, we can achieve optimal performance. 
		If there is no  $ \textit{L}_{ano}$ loss, the performance of virtual face recognition will decrease, which results poor synchronism. $\textit{L}_{div}$ plays a crucial role in the diversity of virtual faces. Without $\textit{L}_{dif}$, the virtual faces of different identities generated under the same key may belong to the same identity, leading to identity collisions.

		\subsubsection{Each Loss of KVFA}
		Similarly, we remove each loss component from the overall training objective of KVFA and summarize the recognition accuracy of the KVFA in Table \ref{tab:9}. When either $\textit{L}_{tot1}$ or  $\textit{L}_{tot2}$ is removed, the recognition accuracy of the KVFA module will significantly decrease. When $\textit{L}_{tot1} $ is removed, KVFA will lose its ability to prevent misidentification, faces of different identities will be recognized as the same identity. When $\textit{L}_{tot2}$ is removed, KVFA will lose its ability to prevent false rejection and faces with the same identity will be recognized as different identities. When both losses participate in the training of KVFA, the model achieves the best recognition performance.
		
		\subsubsection{Threshold Setting for KVFA}
\begin{figure}[!t]
			\centering 
			\includegraphics[width=\linewidth]{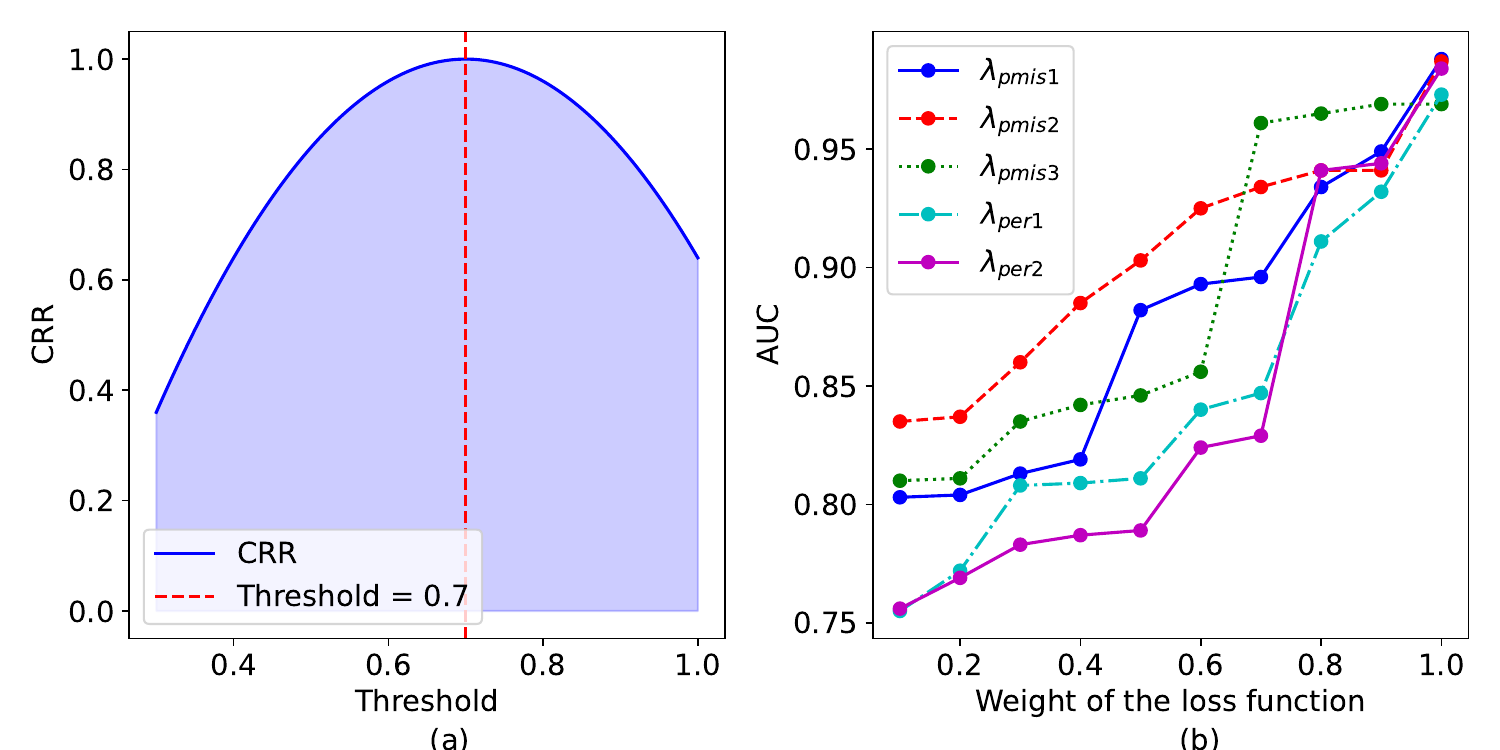}
			\caption{Parameter setting for KVFA. (a) Threshold setting, (b) weight setting for the loss functions.}
			\label{fig_9}    
		\end{figure}

		To evaluate the impact of different threshold setting on identity authentication, we use the validation set of LFW to see how CRR changes under different thresholds, the results of which are shown in Figure \ref{fig_9}(a). It can be seen that we can obtain the highest CRR when the threshold is set as 0.7.

		\subsubsection{Weight Setting for the Loss Functions of KVFA and HPVFG} 
		We evaluate the performance of our method on the validation set of LFW by varying the weights of different loss components in the total losses of KVFA and HPVFG. For the weight setting of the loss function of the KVFA, we gradually change each weight (i.e., $\lambda_{pmis1} $, $\lambda_{pmis2} $, $\lambda_{pmis3} $, $\lambda_{per1} $ and $\lambda_{per2} $) from 0.1 to 1 with the rest four weights being 1, to see how the AUC changes, the results of which are shown in Figure \ref{fig_9}(b). It can be seen that we can achieve the best performance by setting all the weights as 1. 
		
\begin{figure}[!t]
			\centering 
			\includegraphics[width=\linewidth]{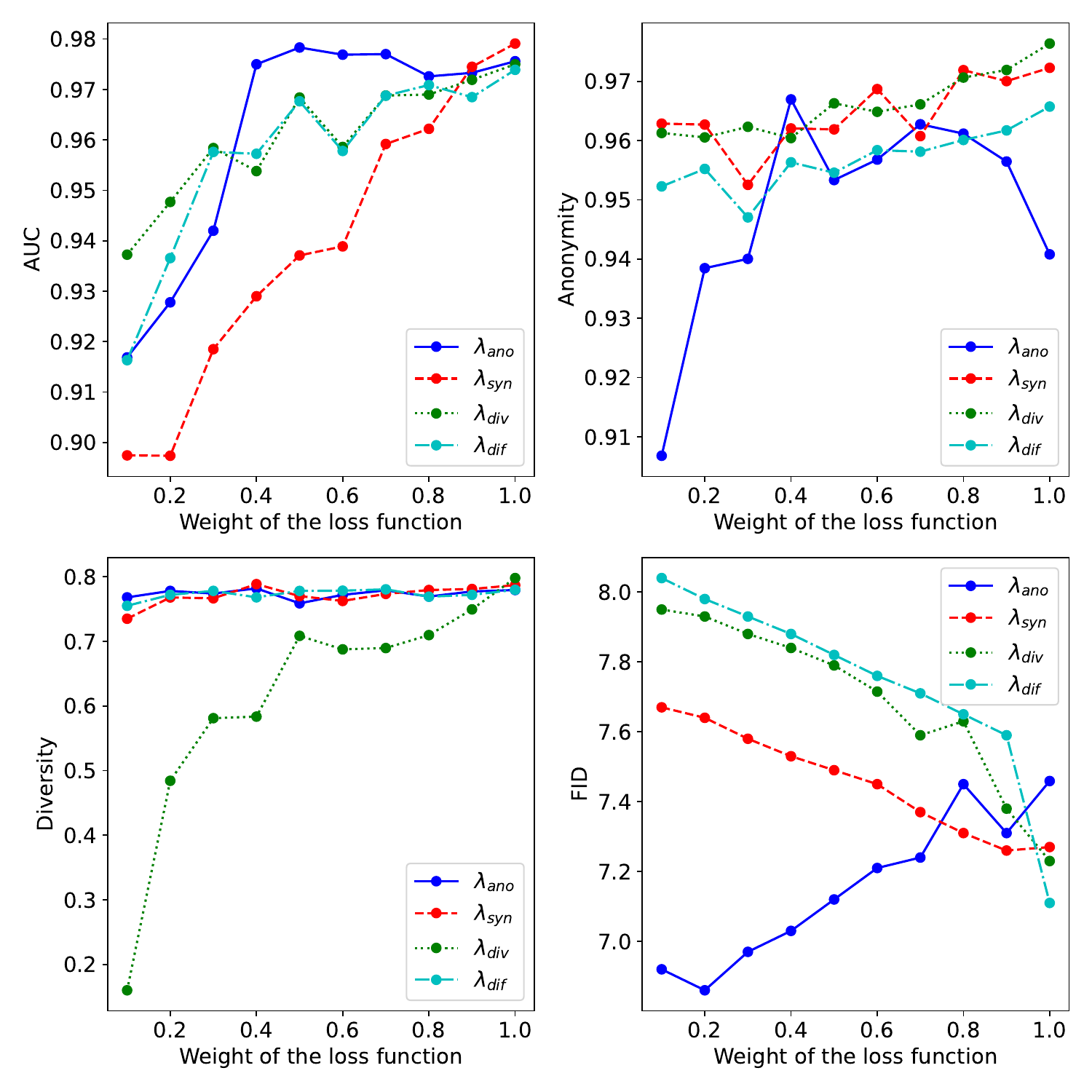}
			\caption{Weight setting for the loss functions of HPVFG.}
			\label{fig_11}    
		\end{figure}

		For the same token, we gradually change each weight (i.e.,  $ \lambda_{ano}$, $ \lambda_{syn} $, $ \lambda_{div} $ and $ \lambda_{dif} $) from 0.1 to 1 in the total loss of HPVFG to see how the performance of the virtual faces would be changed. Figure \ref{fig_11} plots the AUC, anonymity, diversity and FID of the virtual faces under different weight setting. It can be seen that, when we set the $ \lambda_{ano} $ as 0.4, and the other weights as 1, we can achieve a good balance among the AUC, anonymity, diversity and FID for the virtual faces.

		\section{Security Analysis and DISCUSSIONS}
		\subsection{Privacy Leakage}
		
		We note that successful extraction of the original identity from the virtual face requires two conditions: obtaining the user key that controls the generation of the virtual face and the authentication server VFAS. Therefore, privacy breaches may mainly occur during the face image upload stage and the virtual face authentication stage.
		
		\subsubsection{Face Image Upload Stage}
		U sends a key and the original face image to FAS, which generates a virtual face image based on the user key. As an honest server, FAS will automatically delete the original face image and key after the generation of virtual faces. At this stage, the adversary can only obtain the virtual face whose virtual identity is different from the original identity. 
		
		\subsubsection{Virtual Face Authentication Stage} 
		In the process of verifying the original identity of the virtual faces, the adversary needs to hold the correct key and obtains the usage permission of the KVFA model to perform the face authentication. We here discuss two scenarios where only the key is leaked or the KVFA model is leaked.
		
		$\bullet$ \textit{User Key Leakage:}  When the adversary holds the correct key but does not have access to the KVFA model. The adversary may input the virtual face and key into the virtual face generation model HPVFG to recover the original face. As shown in Figure \ref{fig_8}, the first row is the original face, the second row is the virtual face, and the third row is the recovered face. It can be observed that the recovered face is visually different from the original face.
		It can also be seen from Scenario 1 in Table \ref{tab:7} that the similarity between the recovered and the original faces are very low.
		Therefore, from the perspective of visual perception and machine recognition, even if the key is leaked, the face authentication cannot be performed correctly.

		$\bullet$ \textit{VFAS Access Violation:}  When the adversary obtains access to VFAS but does not obtain the correct user key, they are unable to obtain the original identity from the virtual faces. Please refer to Scenario 3 in Section \Rmnum{5}-C-2) for details.  
		
		\subsection{Security Analysis of Key}
		We consider the proposed framework to be secure if the adversary cannot obtain the original identity of our virtual face.
		
		\begin{itemize}
			\item {\textbf{Key Generation:}} The generation of keys uses a secure random number generator, KeyGen, to ensure that each key is unique. In addition, the key generation is done on the user side to prevent it from being obtained by the  adversaries.
			\item {\textbf{Key Storage:}} The key is only stored on the user side. After the virtual face generation, FAS will delete the user key.
			\item {\textbf{The Key Length:}} The longer the key is, the harder it is to be guess. We discuss below the impact of key length on the performance of the virtual faces.
			
		\end{itemize}
		
		\begin{table}[t]
			\caption{ The impact of the key length on the virtual faces.}
			\center
			\label{tab:10}
			\begin{tabular}{cccc}\toprule
				key length& Anonymization &FID   \\ \midrule
				8 bits	&0.962&7.29 \\
				128 bits	&0.956 &\textbf{6.95}\\
				256 bits &\textbf{0.964}& 7.02
				\\	\bottomrule
			\end{tabular}
		\end{table}
		
		\begin{table}[t]
			\caption{ Fault tolerance of different key length.}
			\center
			\label{tab:11}
			\begin{tabular}{cccccc}\toprule
				\multicolumn{1}{c}{\multirow{2}{*}{	key length}} & \multicolumn{5}{c}{ Key error bit}\\ 
				\cmidrule(r){2-6} 
				
				& 0 bit  & 1 bit& 3 bits & 5 bits &16 bits\\ \midrule
				8 bits	& 0.811  & 0.708 & 0.622 & 0.492&-- \\ 
				128 bits	&  0.803 & 0.691 & 0.640 & 0.523 &0.522\\
				256 bits & 0.862 & 0.702& 0.603 & 0.595 &0.485\\	\bottomrule
			\end{tabular}
		\end{table}
		\subsubsection{Impact of the Key Length on the Virtual Faces}We select randomly generated 8-bit, 128-bit, and 256-bit keys to generate virtual facesfrom the LFW testing set. Then we quantitatively evaluate the anonymity of the virtual faces. The experimental results are shown in Table \ref{tab:10}. It can be seen that regardless of the key-length, the anonymity of the virtual face is around 0.96, with FID around 7. Among them, the 256-bit key has the best anonymity, and the 128-bit key has the best image quality. Therefore, it can be concluded that different the key-length has a relatively small impact on the anonymity and image quality of the virtual faces.

		\subsubsection{Fault Tolerance of the Keys With Different Length} We further use the above generated virtual faces for identity authentication, where 0 bit, 1bit, 3 bits, 5 bits, and 16 bits of the key are wrong. The feature (i.e., the output of KVFA) similarity between the virtual face  and the original face is given in Table \ref{tab:11}.
		We believe that if the feature similarity between the virtual face and the original face is less than 0.7, they are considered to be not matched to result a failure of identity authentication. From Table \ref{tab:11}, it can be seen that when the length of the key is 8 bits with 1 bit being incorrect, the feature similarity between the original face and the virtual face is 0.708. When the length of the key is 128 bits and 256 bits with 1, 3, and 5 bits being incorrect, the feature similarity between the original face and the virtual face is less than 0.7, which results in a failure of identity authentication.
		
		Therefore, in practical applications, we suggest to set the key length to be over 128 bits to have a small fault tolerance of the incorrect key bits.
		
		\subsection{Discussions}	
		
		\subsubsection{Potential Threats}
		In practical applications, it is difficult for the adversary to obtain the correct key and the virtual face generation/identity authentication model simultaneously. Therefore, the adversary may engage in illegal identity authentication through some other means.
		
		$\bullet$ \textit{Train a Surrogate Model:} The adversary can train a surrogate KVFG model, which can produce similar features for the original face and the virtual face without the use of a key. 
		

		$\bullet$ \textit{Train an Inversion Model:} The adversary can train an inversion model, the process of which is similar to the inverse process of HPVFG. Specifically, the adversary inputs the the virtual faces into this model to output the original faces. 
		By using such a model, the adversary can obtain both the original identity and appearance of the virtual faces.
		
		However, the above two cases are difficult to be launched in practice. Both of them require collecting sufficient original-virtual face pairs corresponding to the same key for training. Such a behavior can be detected by the service provider. Besides, obtaining diverse
		data is challenging as the users are only able to produce original-virtual face pairs with their identities.

		\subsubsection{Limitations of Randomly Selected Keys}
		In Section \Rmnum{4}-B, we discuss the properties of virtual faces in theory, where the key is randomly chosen. One of our goals is to generate a virtual face which is different from the original face visually and statistically (by using ordinary face recognizers) to achieve high anonymity. In practice, however, some edge cases may happen. First of all, the original and virtual faces may be far away in the feature space but have similar visual appearance in the image space, or vise versa. Secondly, the original and the virtual faces may be close in both the feature space and the image space. These will result in a failure of privacy protection. 
		
		To deal with such edge cases, one possible solution is to carry out a similarity check between the original and virtual faces. If the virtual face falls into one of the edge cases, we revoke it and use a different key to generate another virtual face. Such a process can be repeatedly done until we generate a virtual face with satisfied properties.
		%
		%
		%
		%
		%
		\section{Conclusion}\label{conclusion}
		In this paper, we propose a KFAAR framework for identifying-preserving face anonymization. In this framework, the HPVFG module can generate the virtual faces controlled by a key, while the KVFA module accomplishes the goal of extracting the virtual face's original identity without exposing the visual content of the original face.  We propose and incorporate multi-task learning strategies to train for each module.  Experimental results show that our method achieves good performance in generating head posture and facial expression preservation virtual faces with high anonymity, synchronism and diversity, which can be also used to obtain the original identity when the correct key is given.
		
		\section*{Acknowledgment}
		This work was supported by the National Natural Science Foundation of China under Grant 62072295 and 62072114.
		
		
		
		\bibliographystyle{IEEEtranS}
		\bibliography{sample-base}
		
		%
		%
		%

	\end{document}